\documentclass{article}

\usepackage{PRIMEarxiv}

\usepackage[utf8]{inputenc} % allow utf-8 input
\usepackage[T1]{fontenc}    % use 8-bit T1 fonts
\usepackage{hyperref}       % hyperlinks
\usepackage{url}            % simple URL typesetting
\usepackage{booktabs}       % professional-quality tables
\usepackage{amsfonts}       % blackboard math symbols
\usepackage{nicefrac}       % compact symbols for 1/2, etc.
\usepackage{microtype}      % microtypography
\usepackage{lipsum}
\usepackage{fancyhdr}       % header
\usepackage{graphicx}       % graphics
\graphicspath{{media/}}     % organize your images and other figures under media/ folder

\title{A Short Review for Ontology Learning: Stride to Large Language Models Trend
}

\author{
  Rick Du, Huilong An, Keyu Wang\\
  BSH Home Appliances Holding (China) Co., Ltd\\
  \texttt{\{Rick.Du, Huilong.An, Keyu.Wang\}@bshg.com} \\
   \And
  Weidong Liu \\
  Department of Computer Science and Technology, Tsinghua University \\
  \texttt{liuwd@tsinghua.edu.cn} \\
}

\begin{document}
\maketitle

\begin{abstract}
%\lipsum[1]
Ontologies provide formal representation of knowledge shared within Semantic Web applications. Ontology learning involves the construction of ontologies from a given corpus. In the past years, ontology learning has traversed through shallow learning and deep learning methodologies, each offering distinct advantages and limitations in the quest for knowledge extraction and representation. A new trend of these approaches is relying on large language models (LLMs) to enhance ontology learning. This paper gives a review in approaches and challenges of ontology learning. It analyzes the methodologies and limitations of shallow-learning-based and deep-learning-based techniques for ontology learning, and provides comprehensive knowledge for the frontier work of using LLMs to enhance ontology learning. In addition, it proposes several noteworthy future directions for further exploration into the integration of LLMs with ontology learning tasks.
\end{abstract}

% keywords can be removed
%\keywords{First keyword \and Second keyword \and More}

\section{Introduction}

Extraction and organization of meaningful conceptual knowledge have been central to the pursuit of enhancing machine comprehension and reasoning capabilities \cite{ontologyTool}. Ontology learning, a fundamental cornerstone within this domain, is tasked with the extraction, representation, and refinement of structured ontologies that encapsulate the intricacies of various domains \cite{principles4onto}.  

In the past years, ontology learning has traversed through shallow learning and deep learning methodologies, each offering distinct advantages and limitations in the quest for knowledge extraction and representation \cite{dl4ol1}.  Shallow learning techniques, characterized by their simplicity and ease of implementation, have long been the bedrock of ontology learning \cite{olreview}. These methods, albeit effective in certain contexts, often grapple with challenges of scalability and the extraction of nuanced and complex relationships between entities. Conversely, the advent of deep learning techniques has heralded a new era, promising more intricate representations and enhanced discernment of underlying patterns within data. However, deep learning techniques come burdened with their own set of limitations, including the voracious appetite for large volumes of annotated data and computational resources \cite{dl4ol2}.

Amidst this landscape, the emergence of large language models stands as a disruptive force, reshaping the contours of ontology learning \cite{LLM4OL2, 2_BabaeiGiglou2023LLMs4OLLL}. These models, leveraging the prowess of pre-trained language representations, exhibit a remarkable aptitude for understanding semantic nuances, capturing context, and inferring relationships among entities \cite{LLM4OL2, 2_BabaeiGiglou2023LLMs4OLLL, taxonomy1, taxonomy2}. Their applications in ontology learning holds the promise of addressing longstanding challenges by harnessing the inherent linguistic and conceptual understanding embedded within these models.

The purpose of this paper is to give a review in approaches and challenges of ontology learning in LLMs era. It presents the methods and analyzes the limitations  of shallow-learning-based and deep-learning-based techniques, and provides comprehensive knowledge for the current work of using LLMs to enhance ontology learning. In addition, it proposes several noteworthy future directions for further
exploration into the integration of large language models with ontology learning tasks. The rest of this paper is organized as follows: Section \ref{2def} defines ontology, ontology learning, and summarises the challenges of ontology learning. Section \ref{4method} presents the ontology learning approaches based on shallow learning and deep learning, as well as their limitations. Section \ref{5LLM} presents how large language models contributes to ontology learning procedure recently, and
discusses the potential of using large language models to facilitate ontology learning. Section \ref{direction}  proposes several future directions for further exploration into using large language models to enhance ontology learning. Finally, we conclude in Section \ref{7con}.

\section{Ontology Learning}
\label{2def}
\subsection{Ontology}
% For the formal definition of ontology, there are various interpretations and applications of this term, some of which are explored in\cite{ontologyDef}. For clarity in this paper, we adopt the simple formal definition of "ontology" as proposed in\cite{TOO} and elaborated in this section.\\
% \cite{ontologyDefPhi} describe "ontology" as a theoretical framework concerning existence. In the field of Artificial Intelligence, ontology is viewed as a formalized specification of a particular domain's concepts. This includes detailing their relationships, constraints, and axioms, thereby establishing a shared lexicon for knowledge exchange\cite{Gruber1995TowardPF}. Essentially, what needs to be encapsulated in a knowledge-based system is that which exists, aligning these definitions as complementary to each other.\\
In general, an ontology describes formally a domain of discourse. Typically, an ontology consists of terms and the relationships between these terms, where the terms denote important concepts of the domain  \cite{semanticweb}.  An ontology must be formal and machine-readable, allowing it to serve as a shared vocabulary across different applications.  Formally, ontology can be described as following tuple \cite{TOO}:
\begin{center}
    \begin{equation} O = <C,H,R,A> \end{equation}
\end{center}
where $O$ represents ontology, $C$ represents a set of classes (concepts), $H$ represents a set
of hierarchical links between the concepts (taxonomic relations), $R$ represents a set of
conceptual links (non-taxonomic relations), and $A$ represents a set of rules and axioms.

\subsection{Ontology Learning}

% And the process of constructing an ontology involves several critical steps, which are outlined as follows:
% \begin{enumerate}
%     \item \textbf{Define the domain and scope:}Identify the specific domain (e.g., healthcare, finance) and define the scope and determine the level of detail required.
%     \item \textbf{Consider reuse of existing ontologies:} Investigate existing ontologies for potential reuse or extension besides creating an ontology from scratch.
%     \item \textbf{Identify key concepts and relationships:} List major concepts and identify relationships and properties.
%     \item \textbf{Investigate rules and axioms:} Determine or deduce the principles and propositions that characterize the domain's structural attributes.
%     \item \textbf{Create the ontology:} Use an ontology development tool and a formal ontology language.
%     \item \textbf{Evaluation and validation:} Evaluate and validate the ontology with predefined criteria and domain experts.
%     \item \textbf{Iterative refinement:} Continuously refine the ontology based on feedback.
% \end{enumerate}

% \section{Ontology Construction via Learning}
% \label{3ol}

Ontology learning (OL) from text involves the  construction of ontologies from a given corpus of text \cite{ontologyDefRef, pekar2002taxonomy}. According to ontology learning layer cake shown in Figure \ref{fig:layercake} proposed by \cite{eval2}, which is widely held as cornerstone in OL \cite{Browarnik2015OntologyLF}, the process of OL from text can be divided into six sub-tasks as following:

\begin{figure}[h]
\centering
\includegraphics[width=0.8\textwidth]{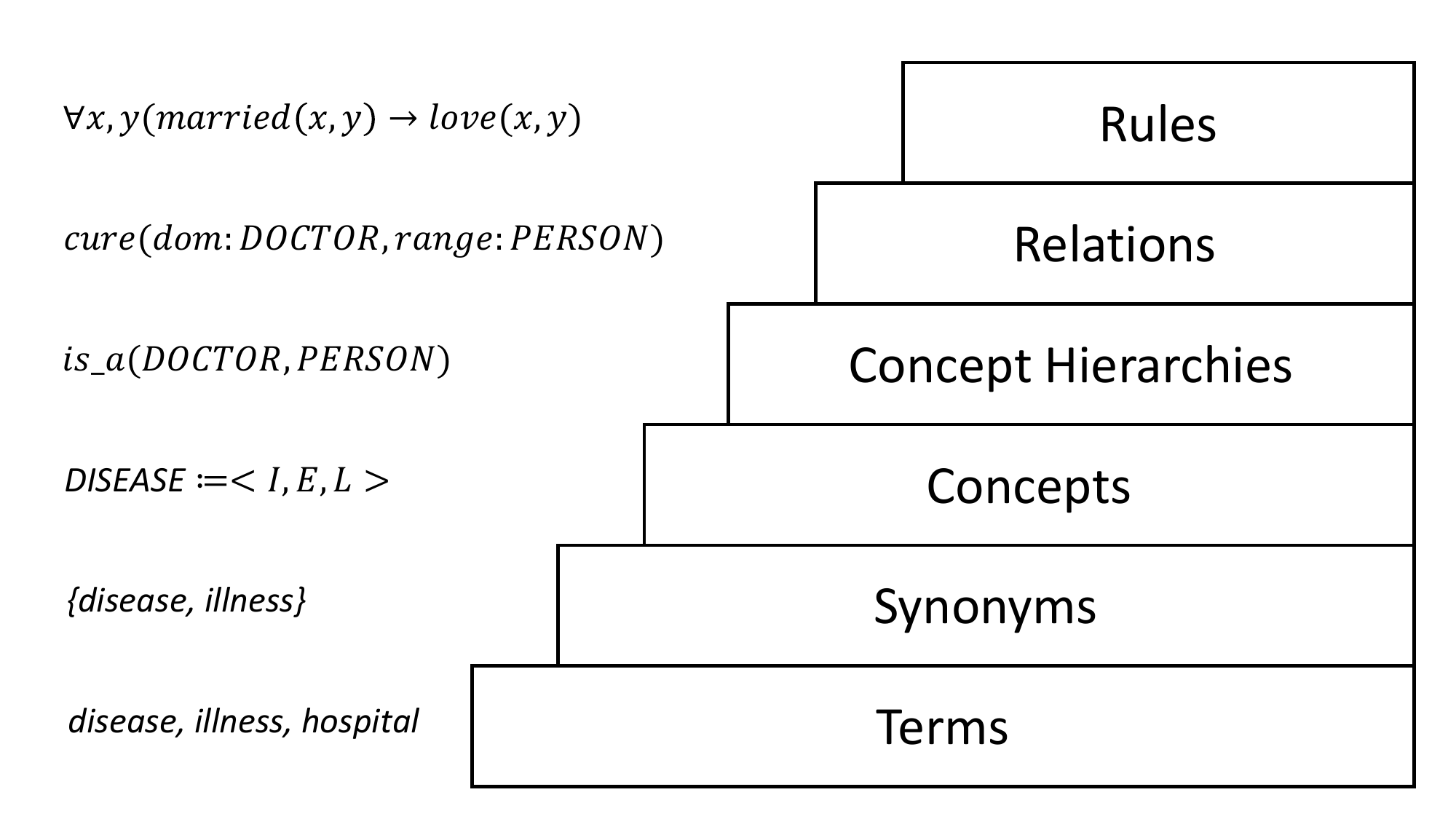}
% 后面重新做
\caption{Ontology Learning Layer Cake \cite{eval2}}
\label{fig:layercake}
\end{figure}

\begin{enumerate}
    \item \textbf{Term extraction:} This initial step involves identifying relevant terms or entities from a given text or dataset. These terms serve as the building blocks for constructing an ontology. 
    \item \textbf{Synonym extraction:} Synonyms are different terms referring to the same concept. In ontology learning, identifying synonyms is crucial for ensuring comprehensive coverage and avoiding redundancy.
    \item \textbf{Concept formation:} Once terms and their synonyms are extracted, the next step is to group them into meaningful concepts or classes. This involves organizing related terms into hierarchies or categories based on their similarities, functionalities, or semantic relations. 
    \item \textbf{Taxonomic relation extraction:} Taxonomic relations establish hierarchical relationships between concepts, defining the "is-a" relationship (e.g., "car" is a "vehicle"). Ontology learning involves identifying and structuring these hierarchical relationships to arrange concepts in a taxonomy or ontology hierarchy. 
    \item \textbf{Non-taxonomic relation extraction:} Unlike taxonomic relations, non-taxonomic relations capture various associations between concepts beyond hierarchical relationships. These relations could be "part-of," "has-property," or other associative connections that enrich the ontology's expressiveness. 
    \item \textbf{Rule or axiom extraction:} Rules or axioms define constraints, dependencies, or logical relationships between entities or concepts in the ontology. Extracting rules or axioms aims to formalize domain knowledge and establish logical constraints within the ontology. 
\end{enumerate}

%According to \cite{ontologyConstruct1,ontologyConstruct2,dl4ol1,ontologyConstruct4,ontologyConstruct5}, ontology construction is an iterative process and could be done in three ways:
Generally, the ontology learning process follows the aforementioned steps. However, it is not uncommon for some ontology learning processes only partially complete the six steps outlined above according to different needs. Ontology learning methods can be roughly divided into the following three categories \cite{ontologyConstruct1,ontologyConstruct2,ontologyConstruct4,ontologyConstruct5}:

\begin{itemize}
\item \textbf{Manual:} Ontologies are developed through a process that heavily relies on human expertise and intervention. Examples are Gene Ontology (GO) \cite{ManualOntologyExample_GO}, WordNet \cite{ManualOntologyExample_wordnet}, SNOMED CT (Systematized Nomenclature of Medicine—Clinical Terms) \cite{ManualOntologyExample_CT}, Cyc \cite{ManualOntologyExample_cyc}, and Foundational Model of Anatomy (FMA) \cite{ManualOntologyExample_fma}.
\item \textbf{Semi-automatic:} The development of ontologies is facilitated and streamlined by integrating automated processes with human input. There are various available tools for such a purpose, like Text2Onto \cite{text2onto2005}, OntoGen \cite{ontogen2008}, and OntoStudio \cite{ontostudio2020}. 
\item \textbf{Fully automatic:} The system takes care of the complete construction, without any manual intervention. While the idea of fully automatic ontology construction is appealing, especially for handling large volumes of data or complex domains, it is worth mentioning that full automatic construction for ontology by a system is still a significant challenge and it is not likely to be possible \cite{autoOntoAI,autoOntoChallenges,autoOntoNLP}.
\end{itemize}

\subsection{Challenges in Ontology Learning}
Ontology learning, despite its advancements, still encounters various challenges. Below is a list highlighting the key aspects that characterize the primary challenges in ontology learning:

\textbf{Labor intensiveness:} Ontology construction often involves significant manual effort. Identifying, extracting, and structuring knowledge from diverse sources demands extensive human intervention. This labor-intensive process can be time-consuming and resource-intensive, hindering the scalability and efficiency of ontology development \cite{Browarnik2015OntologyLF,Challenge1,Challenge2, eval2}.

\textbf{Axiom formulation:} Formulating precise axioms or rules that accurately represent domain knowledge poses a challenge. Balancing expressiveness with computational efficiency is crucial. Axioms must be meaningful and precise to contribute effectively to the ontology's utility. This demands specialized expertise and often involves iterative refinement \cite{Mishra2015ASO,Challenge1,Challenge2}.

%\textbf{Relation and axiom discovery:} Accurately defining both hierarchical and non-hierarchical relations is critical and can be difficult, especially in complex domains\cite{Browarnik2015OntologyLF,Challenge1,Challenge2, eval2,eval1}, and discovering axioms that accurately represent domain knowledge requires a careful balance between expressiveness and computational efficiency which is still in the initial stage \cite{Mishra2015ASO,Challenge1,Challenge2}.

%\textbf{Learning and formulating axioms:} Developing axioms that accurately represent domain knowledge requires a careful balance between expressiveness and computational efficiency. Axioms must be precise and meaningful to effectively contribute to the ontology's utility. However, axioms discovery is still in the initial stage and requires intensive work\cite{Mishra2015ASO,Challenge1,Challenge2}.

\textbf{Domain-specific knowledge acquisition:} Acquiring and representing domain-specific knowledge within the ontology is challenging. Understanding and capturing intricate domain nuances, concepts, and relationships require expert domain knowledge. Incorporating evolving or specialized domain terminologies into the ontology accurately is complex \cite{acqui1, acqui2}.

\textbf{Dynamic environments:} Adapting ontologies to dynamic or evolving environments is challenging. Ensuring ontology coherence and consistency while accommodating changes in domain concepts, terminologies, or relationships demands continuous updates and version control mechanisms \cite{evolu1, migra1}.

\textbf{Ambiguity and uncertainty:} Dealing with ambiguous terms, uncertain knowledge representations, or conflicting information within data sources presents challenges. Resolving ambiguity and handling uncertain or conflicting data affect the ontology's accuracy and reliability \cite{Challenge1,Challenge2}.

\textbf{Scalability:} Ontology learning must accommodate large-scale data and knowledge sources while maintaining computational efficiency. Scaling ontology construction methods to handle substantial volumes of data without sacrificing accuracy remains a significant challenge \cite{scale1, scale2}.

\textbf{Heterogeneity of data:} Integrating heterogeneous data from diverse sources, each with different structures, formats, and semantics, presents challenges. Aligning and reconciling conflicting data representations and resolving semantic mismatches is crucial for creating coherent and comprehensive ontologies \cite{eval2,Mishra2015ASO, Challenge1,Challenge2}. 

\textbf{Evaluation and validation:} Properly evaluating ontologies for accuracy, completeness, and usability is complex. Defining reliable evaluation metrics, validation methods, and assessing ontology quality pose challenges due to the subjective nature of evaluating knowledge representations 
 \cite{Challenge1,Challenge2,eval1,eval2}.

\section{Ontology Learning Approaches}
\label{4method}
\subsection{Shallow-learning-based Approaches}
Before the rise of deep learning, shallow learning methods grounded in traditional machine learning and classical neural networks was predominant in ontology learning tasks such as term extraction, concept formation, taxonomy discovery, non-taxonomic relation extraction, and axiom extraction \cite{dl4ol1}. These techniques mainly fall into three categories \cite{dl4ol1,Asim2018ASO}: 
\begin{itemize}
    \item \textbf{Linguistics-based approaches.} Linguistic techniques are based on characteristics of language,  such as pattern-based extraction \cite{Morin_pattern}, POS tagging and sentence parsing \cite{AbneyPOS}, syntactic structure analysis and dependency structure analysis \cite{NivreDependency,Gamallo2002Depend} and etc. 
    \item \textbf{Statistics-based approaches.} Statistical techniques are based on statistics of the underlying corpora. Typical methods include co-occurrence analysis \cite{BudanitskyCooccur}, association rules \cite{MaedcheAsso}, heuristic and conceptual clustering \cite{Faure1998,Faure2000}, contrastive analysis \cite{Navigli2002Contrast}, latent semantic analysis (LSA) \cite{Rani2017LSA, Landauer1998LSA}, term subsumption \cite{Fotzo2004LearningR} and etc. 
    \item \textbf{Logic-based approaches.} Logic-based techniques are based on formal logic and reasoning. Typical methods include inductive logic programming \cite{ZelleILP}, logical inference \cite{ShamsfardLogic} and etc.
\end{itemize}
%This section will first summarize various methods from the past decade applied to Ontology Learning tasks as table \ref{table:1}, and later we list some of limitations posed by shallow learning methods.

Despite its simplicity, speed, interpretability, less data intensive and ease of implementation, shallow learning for ontology learning have several drawbacks and limitations:
\begin{itemize}
    \item \textbf{Limited capacity for autonomous inference: } Shallow learning methods typically lack the ability to autonomously infer new relations or axioms. They rely on predefined patterns or rules and struggle to extrapolate beyond the explicitly provided data patterns. Consequently, these systems suffer from poor recall, meaning they may miss or fail to capture important relationships or concepts not explicitly present in the training data \cite{dl4ol1}.
    \item \textbf{Small, domain-specific datasets:} Shallow-learning-based systems often operate with limited datasets, especially in domain-specific contexts. These datasets might not adequately represent the richness and complexity of the entire domain. As a result, the resulting ontologies may lack depth, completeness, or accuracy, failing to capture the nuanced relationships and concepts within the domain.
    %\item \textbf{Inadequate representation of domain knowledge:} Due to the limited scope of data and shallow inference capabilities, ontologies constructed through shallow learning may not accurately represent the diverse, complex, and evolving nature of domain knowledge. They might overlook subtle relationships, variations, or evolving trends within the domain, leading to less comprehensive ontologies.
    \item \textbf{Scalability concerns:} Shallow learning approaches might struggle to scale effectively when faced with large, diverse datasets. They might not handle the complexities of diverse data sources, leading to difficulties in integrating and reconciling disparate information into a cohesive ontology representation \cite{Andleeb2015FromTM}.
    \item \textbf{Dependency on human intervention:} Shallow-learning-based systems may heavily rely on human intervention. They often require manual input and supervision at various stages. This dependency on human expertise makes the process time-consuming, resource-intensive, and less scalable \cite{dl4ol1}.
\end{itemize}

Although shallow learning has limitations in integration within ontology learning, it provides a valuable and easily deployable framework for the initial automatic creation of ontologies. This serves as a precursor for subsequent applications of more complex deep learning-based methods or expert-driven refinements.

\begin{table}[h!]
\centering
\label{table:1}
\begin{tabular}{|c|c|}
\hline
\textbf{OL Approaches} & \textbf{OL Tasks} \\ \hline
\multicolumn{2}{|c|}{\textbf{Linguistics Based}} \\ \hline
Pattern-based extraction \cite{Morin_pattern} & Concept hierarchy \& Relation discovery \\ \hline
POS tagging \& Sentence parsing \cite{AbneyPOS} & Term extraction \\ \hline
Syntactic structure analysis \& & Term extraction, Concept \\
Dependency structure analysis \cite{NivreDependency,Gamallo2002Depend} & hierarchy \& Relation discovery \\ \hline
\multicolumn{2}{|c|}{\textbf{Statistics Based}} \\ \hline
Co-occurrence analysis \cite{BudanitskyCooccur} & Term extraction \& Concept formation \\ \hline
Association rules \cite{MaedcheAsso} & Relation discovery \\ \hline
Heuristic/conceptual clustering \cite{Faure1998,Faure2000} & Synonym discovery, Concept \\
& formation \& Concept hierarchy \\ \hline
Contrastive analysis \cite{Navigli2002Contrast} & Term extraction \\ \hline
Latent semantic analysis \cite{Rani2017LSA, Landauer1998LSA} & Concept formation\\ \hline
Term subsumption \cite{Fotzo2004LearningR} & Concept hierarchy\\ \hline
\multicolumn{2}{|c|}{\textbf{Logic Based}} \\ \hline
Inductive logic programming \cite{ZelleILP} & Axiom extraction \\ \hline
Logical inference \cite{ShamsfardLogic} & Concept hierarchy \& Relation discovery \\ \hline
%\multicolumn{2}{c}{\textbf{Deep Learning Based}} \\ \hline
%& \\ \hline
%& \\ \hline
%& \\ \hline
%& \\ \hline
%& \\ \hline
%& \\ \hline
\end{tabular}
\caption{Shallow-learning-based ontology learning approaches and their corresponding tasks.}
\end{table}

\subsection{Deep-learning-based Approaches}
It is noteworthy that in the last few years, deep learning techniques have achieved significant performances in various natural language processing tasks such as machine translation \cite{DLapp1} and sentiment analysis \cite{DLapp2}. Extensive works have demonstrated that deeper analysis excelled in understanding texts compared to shallow learning \cite{dl4text1,dl4text2, dl4text3}. Specifically, they have been applied to ontology learning procedure such as concept extraction and relation extraction \cite{dl4ol1}.

\paragraph{Concept extraction.} Typically, ontology learning begins by recognizing and extracting terms and their synonyms from corpora, and then these terms and synonyms are combined to form concepts \cite{dl4ol2}. Named entity recognition (NER) identifies essential terms that form the basis for defining concepts and relationships in the ontology. \cite{NERreview1, NERreview2, NERreview3} provide comprehensive reviews on deep learning techniques for NER. 
Some studies have dedicated their attention to term extraction in a specific domain, which can enhance domain ontology learning. For instance,  \cite{IndoLanNER} studied the named entity recognition in Indonesian language for concept extraction in building an ontology. This study provided an end-to-end system based on BiLSTM, which can be used for part-of-speech tagging and named entity recognition without using additional tools for part-of-speech tagging. \cite{AgriNER} proposed a two-stage process using a semantic-based deep learning approach to extract terms in the agricultural domain. In the first stage, the semantic-based method was employed to detect agricultural entities and semi-automatically construct a labeled agricultural entity corpus. In the second stage, \cite{AgriNER} utilized deep learning techniques to identify agricultural entities from pure texts. Further, \cite{KGANER} introduced a task-based approach using NLP techniques for domain-specific information extraction, including a bi-LSTM-CRF model for entity extraction, attention-based Semantic Role Labeling, and an automated verb-based relationship extractor.

Once the domain-specific terms are acquired, the next step is to identify synonyms among these terms, leading to the formation of synsets.   \cite{ESN} built distributional word embeddings using Word2Vec \cite{word2vec} and then used the induced word embeddings as an input to train a feedforward neural network using annotated dataset to distinguish between synonyms and other semantically related words. \cite{ASDKB} introduced DPE, a novel framework that seamlessly combines distributional features from corpus-level statistics with textual patterns from local contexts for the automatic discovery of synonyms using a knowledge base.  SynSetExpan \cite{synsetexpan} enables two tasks, synonym discovery and entity set expansion to mutually enhance each other, using a synonym discovery model to include popular entities' infrequent synonyms into the set and a set expansion model to determine whether an entity belongs to a semantic class.  \cite{MES} proposed a framework with two novel modules to mine entity synsets from raw corpus: a set-instance classifier that jointly determines how to represent an entity synset and whether to include a new term, and an efficient set generation algorithm that applies the learned classifier to discover new synsets.

Concept formation follows term extraction and synonym identification. While some methods treat terms as concepts, others cluster similar terms to create concepts \cite{TOO, OLSpan}. Deep learning models, such as BiLSTMs and USE, have shown effectiveness for concept formation \cite{OntoEnricher}.

\paragraph{Relation extraction.} In an ontology, relations are crucial to express relationships between concepts. An effective way to automatically acquire this knowledge, called Relation Extraction (RE), plays a significant role in ontology learning. So far, numerous studies have been conducted on RE, with technologies based on deep neural networks (DNNs) becoming the mainstream direction of this research.

Relations in ontologies encompass hierarchical and non-hierarchical categories. \cite{IsAREsurvey} gave a short review on taxonomy learning from text in terms of issues, resources and recent advances. Despite it discussed the limited success of deep learning paradigms for taxonomy induction since designing a single objective for neural networks to optimize is difficult, \cite{IsAREsurvey} suggested researching how to utilize deep learning for taxonomy induction is worthwhile in the future. \cite{HRE} proposes a novel hierarchical attention scheme to incorporate hierarchical information for distant supervision. The multiple layers of the hierarchical attention scheme provide coarser-to-fine granularity to better identify valid instances, especially for extracting long-tail relations. 

More research work concentrated on non-hierarchical relation extraction. Han et al. \cite{M} classified RE methods into pattern extraction, statistical extraction, and neural extraction, with a focus on neural relation extraction (NRE) using deep learning models like CNNs, RNNs, GNNs, and attention-based networks. Apart from supervised methods like CNN, RNN, LSTM and pre-trained models such as BERT , Aydar M, et al \cite{NREsurvey} further discussed the recent advances of distant supervised methods and few-shot methods on relation extraction. \cite{nre1} used Semantic and Thematic Graph Generation Process  for automatic relationship construction in domain  ontology engineering.

\paragraph{Limitations.} While deep learning holds promise for ontology learning, several drawbacks and limitations currently hinder its widespread application:

\begin{itemize}
    \item \textbf{Data requirements and labeling:} In ontology learning, acquiring annotated ontologies or extensive labeled data for training deep learning models might be challenging due to the specificity and complexity of domain knowledge.

    \item \textbf{Semantic understanding and context sensitivity:}  Ontologies require precise representations of concepts and relations, and deep learning models might face challenges in capturing these intricate semantics accurately. Deep learning models might struggle with capturing subtle semantic nuances or understanding contextual variations within domain-specific terminology.

   % \item \textbf{Computational Resource Requirements:} Training deep learning models, especially complex neural networks, demands significant computational resources, including high-performance hardware and substantial processing power. These computational requirements might be prohibitive for many ontology learning tasks, particularly in resource-constrained environments.
    \item \textbf{Domain adaptation and transfer learning:} Ontology learning often involves diverse domains with varying characteristics. Adapting deep learning models to new domains and utilizing transfer learning techniques are challenging, especially when dealing with domain-specific nuances and specialized terminologies.
    
    %\item \textbf{Domain specificity and generalization:}  Constructing an ontology that encompasses various domains without bias remains a challenge. Deep learning models might struggle with generalization across diverse domains or fail to capture domain-specific nuances adequately.

    \item \textbf{Expertise and resource requirements:}  Building and fine-tuning deep learning models require specialized expertise in machine learning, neural networks, and computational resources. This can limit accessibility and practical implementation for non-experts when applying to different domain ontologies.

\end{itemize} 

The trend suggests a movement towards mitigating the limitations of deep learning for ontology learning by combining it with other techniques, focusing on  addressing labeled data scarcity, domain adaptation and semantic understanding. Leveraging pre-trained large language models and few-shot learning techniques could mitigate the data dependency issue, domain adaption and computational resource challenges, enabling ontology construction with smaller datasets and lower computational requirements.

\section{LLM era: How It Contributes to Ontology Learning?} 

\label{5LLM}
\subsection{Development of Large Language Models} 
The emergence of large language models (LLMs) has revolutionized the field of natural language processing, propelling the boundaries of artificial intelligence by enabling machines to understand and generate human-like text \cite{5_Wang2023ASO}. These models, built upon deep learning architectures, have undergone a remarkable evolution, fostering unprecedented advancements in various applications such as text generation \cite{llmapp1}, translation \cite{llmapp2}, sentiment analysis \cite{llmapp3}, and information retrieval \cite{llmapp4}.

The genesis of large language models can be traced back to the foundations of neural networks and early attempts at language representation. However, the groundbreaking transformations in this domain primarily began with the development of the Transformer architecture by Vaswani et al. \cite{transformer} in 2017. This architecture addressed long-range dependencies in sequences more effectively than recurrent neural networks (RNNs) and convolutional neural networks (CNNs) by employing attention mechanisms.

The Transformer architecture served as a catalyst for the creation of several pioneering large language models. Notably, the introduction of OpenAI's GPT (Generative Pre-trained Transformer) series marked a watershed moment in the evolution of language models \cite{gpt1}. GPT-1 demonstrated remarkable capabilities in natural language understanding and generation by utilizing unsupervised pre-training on vast corpora, followed by fine-tuning on specific downstream tasks \cite{gpt1}. Subsequent iterations of the GPT series, including GPT-2 \cite{gpt2}, GPT-3 \cite{gpt3}, and their variants, escalated the scale and performance of language models by leveraging increasingly larger datasets and more intricate architectures. GPT-3, released in 2020, with its staggering 175 billion parameters, represented a paradigm shift in the size and capabilities of language models \cite{gpt3}. Its sheer size endowed it with a remarkable aptitude for various language tasks, exhibiting a high degree of flexibility and adaptability with minimal task-specific fine-tuning.

Apart from OpenAI's contributions, other prominent models such as BERT (Bidirectional Encoder Representations from Transformers) proposed by Google and its variations have also significantly influenced the landscape of large language models \cite{bert}. BERT's innovation lies in bidirectional pre-training, enabling the model to capture contextual information bidirectionally, thereby enhancing its understanding of language nuances \cite{bert}.

%\begin{figure}[h]
%\centering
%\includegraphics[width=1.0\textwidth]{images/LLMs.png}
% 后面重新做
%\caption{\textbf{(left)} Transformer architecture %\cite{transformer}. \textbf{(upper right)} GPT architecture \cite{gpt1}. \textbf{(lower right)} BERT architecture \cite{bert}. }
%\label{fig:LLMs}
%\end{figure}

The progression in natural language processing is primarily evident in the sophisticated text interpretation and production abilities of these models. LLMs like GPT-4 demonstrate nuanced understanding of context, subtlety, and complexity in language, generating coherent and contextually relevant text across diverse topics \cite{GPT4_OpenAI2023GPT4TR}. The advancements have been quantified through various benchmarks and metrics in the field, where LLMs have consistently raised the bar for machine understanding of syntax, semantics, and pragmatic language aspects \cite{9_Bubeck2023SparksOA}. These capabilities have transformed sectors reliant on natural language, enabling automated, high-fidelity text generation, and deepening human-AI interaction. Through their extensive pre-training, LLMs have acquired a breadth of linguistic knowledge, making them highly versatile in text-based task performance, from simplifying intricate scientific material to crafting creative literary works \cite{10_Naveed2023ACO}. This represents a significant leap forward in bridging the gap between human-like language processing and artificial intelligence.

The trajectory of large language models has seen an ever-accelerating pace of innovation, with ongoing research focusing on enhancing model efficiency, interpretability, and reducing biases.
Recently, large language models have been applied to various ontology tasks such as ontology matching \cite{bertmap} and inconsistency handling \cite{reasoning}. Despite the fact that currently there is no research explicitly training LLM for ontology learning, there are empirical attempts to verify if LLMs are suitable or superior for ontology learning tasks, with focus on term typing, taxonomy building and non-taxonomic relations discovery and so on. 

\subsection{Concept Extraction}
% Term typing means a generalized type is discovered (automatically) for a lexical term\cite{dl4ol1}. It is a crucial phase in ontology learning, involves categorizing lexical terms into distinct conceptual types, which is the foundation of building structured, meaningful ontologies in various domains\cite{dl4ol1}. it is the center to understanding and organizing domain-specific knowledge and it facilitates automated reasoning and data integration across systems.

Using sophisticated algorithms and neural network architectures, large language models are trained on extensive textual corpora to understand language semantics, context, and intricate patterns. This understanding enables them to discern and extract meaningful concepts from vast amounts of unstructured text with remarkable accuracy and efficiency. There have been extensive researches using pre-trained models, such as BERT, to identify named entities, which is the basis of forming concepts \cite{ner1, ner2, ner3}. In this section, we focus on concept extraction by using large language models in ontology learning procedure.

%\begin{figure}[h]
%\centering
%\includegraphics[width=0.7\textwidth]{images/oellm.png}
%\caption{Examples of the template to convert Natural Language sentences into OWL Functional Syntax \cite{LLM4OL2}.}
%\label{fig:oellm}
%\end{figure}

\cite{LLM4OL2} fine-tuned a GPT-3 model to convert natural language sentences into OWL Functional Syntax and employed objective and concise examples to fine-tune the model regarding: instances, concepts, as well as class subsumption, domain and range of relations, object properties relationships, disjoint classes, complements, cardinality restrictions.  \cite{2_BabaeiGiglou2023LLMs4OLLL} shows that LLMs can leverage their extensive pre-trained knowledge to categorize terms. In particular, LLMs undertake term typing by discerning the contextual placement of a term within a given domain. This can be facilitated through well-crafted prompts in a close or prefix style. A term is presented to the model within a sentence, followed by a [MASK] token indicating the type to be predicted. The model, drawing upon its extensive training, fills in this blank with an appropriate type, effectively classifying the term, with vast data sources. Based on the characteristics of these sources, 8 different prompts have been designed to conduct zero-shot testing on multiple LLMs.  In comparison to traditional term typing methods which often rely on manual categorization or simpler NLP techniques, LLMs show the potential in terms of speed, scalability, and contextual accuracy.

\subsection{Relation Extraction}
% Taxonomy discovery is a subtask of ontology learning that involves the automatic discovery of hierarchical relationships especially “is-a” between concepts or entities in a given knowledge source [cite 6 layers of Ontology learning: 1]. It aims to identify the broader categories or classes that individual concepts or entities belong to, and to organize them into a hierarchical structure. This is an important task in many applications, such as information retrieval, natural language processing, and knowledge management, as it enables more efficient and accurate processing of large amounts of data. 
\paragraph{Hierarchical relation.} Several studies have indicated that the utilization of large language models  for facilitating the identification of taxonomy significantly mitigates the need for manual intervention. \cite{taxonomy1} proposed an approach for taxonomy discovery applying pretrained language models composed of two modules, one that predicts parenthood relations and another that reconciles those predictions into trees. The parenthood prediction module produces likelihood scores for each potential parent-child pair, creating a graph of parent-child relation scores, while the tree reconciliation module treats the task as a graph optimization problem and outputs the maximum spanning tree of this graph. \cite{taxonomy2} analyzed zero-shot taxonomy learning methods which are based on distilling
knowledge from language models via prompting and sentence scoring, and found that these methods outperform some supervised strategies and are competitive with the current state-of-the-art under adequate conditions. Further, \cite{taxonomy3} conducted a systematic comparison
between the prompting and fine-tuning approaches for taxonomy discovery, revealing that  the prompting approach outperforms
fine-tuning-based approaches while taxonomies generated by the fine-tuning approach can be easily post-processed to satisfy all the constraints. To facilitate automatic taxonomy induction, \cite{col} proposes Chain-of-Layer,  an in-context learning framework designed to induct taxonomies from a given set of entities, which breaks down the task into selecting relevant candidate entities in each layer and gradually building the taxonomy from top to bottom. 

Some studies have achieved good results in the process of ontology learning by using large language models to build taxonomy. \cite{2_BabaeiGiglou2023LLMs4OLLL} evaluated the ability of LLMs to automatically discover the taxonomy of a given knowledge source.  4 top-down taxonomy templates for super-class prediction and 4 bottom-up prompt templates for subclass prediction are defined for LLMs.  The results show that LLMs can effectively discover type taxonomies from different knowledge sources.
In addition, the performances of LLMs are compared with that of traditional ontology learning methods, such as lexico-syntactic pattern mining and clustering, showing that LLMs outperform these methods in most cases. \cite{3_Funk2023TowardsOC} defined a method for automatically constructing a concept hierarchy for a given domain by querying a LLM. Based on OpenAI's GPT 3.5, \cite{3_Funk2023TowardsOC} was applied to various domains such as Animals, Drinks, Music, and Plants. The basic idea is to crawl the hierarchy by repeatedly asking the LLM to provide relevant subconcepts of the given concepts that are already in the hierarchy and used an established traversal algorithm to place the new concepts. Further, the LLM was asked to provide a textual description of each concept that they made available for inspection. 

%\begin{figure}[h]
%\centering
%\includegraphics[width=1.0\textwidth]{images/taxonomy_building1.png}
%\caption{Concept Hierarchy Construction %\cite{3_Funk2023TowardsOC}}
%\label{fig:taxonomy}
%\end{figure}

In conclusion, automatically constructing a concept hierarchy for a given domain by querying a large language model like OpenAI's GPT 3.5 can be a valuable investigation domain for constructing concept hierarchies. It is also observed that the performances of LLMs vary depending on the models and the knowledge sources, and the results  can be further improved by adding interaction with a human domain expert. 

\paragraph{Non-hierarchical relation.} Non-hierarchical relations
% discovery is a subtask of ontology learning that involves the automatic discovery of semantic relationships between concepts or entities in a given knowledge source that are not hierarchical in nature [cite ontology learning]. These relationships 
can be of various types, such as part-whole relationships, causal relationships, or associative relationships. 
% Non-taxonomic relationship discovery is to identify these relationships and to organize them into a structured representation, such as an ontology or a knowledge graph. 
Despite the fact that a lot of work has been conducted to extract relations applying large language models \cite{re1,re2,re3,re4}, the work of non-taxonomy extraction using large language models in ontology learning process is still at an initial stage.

\cite{nontax} used BERT to predicted relationships between phrases. Though it mainly applied for generating a taxonomic structure consisting of a hypernym–hyponym relationship from the extracted phrase set,  it could also create non-taxonomic relationships between phrases according to the intended use of the
ontology. \cite{2_BabaeiGiglou2023LLMs4OLLL} defined a test by involving selecting a set of non-taxonomic relations from a given knowledge source, and creating a testing dataset comprising all pairs of types for each relation. The LLMs were then trained on this dataset and evaluated using standard metrics such as precision, recall, and F1-score. UMLS \cite{UMLS} was used as the knowledge source, which contained over 2 million biomedical concepts and their relations. In the defined test, 53 non-taxonomic relations from the UMLS are selected. The results of test showed that LLMs could effectively discover non-taxonomic relations between types from the UMLS knowledge source. It shows that LLMs has the potential to effectively capture the semantic relationships between types in a given knowledge source, and that the performance of LLMs varies depending on the models and the types of non-taxonomic relation.  

\subsection{LLMs Empowered Tools for Ontology Development}
Recently, several tools have been developed to facilitate the application of deep learning and even large language models to ontology learning. \cite{4_He2023DeepOntoAP} has developed a tool named DeepOnto aiming to build up a Python library to (semi-)automatically facilitate ontology engineering with deep learning, which can be used for ontology learning. The library relies on Python programming in synergy with deep learning methodologies, with a particular emphasis on pre-trained language models. For ease of use, DeepOnto has encapsulated basic ontology processing functions by bridging up Python libraries with Java with OWL API, and implemented several essential components such as reasoning, verbalisation, pruning and projection. It incorporates systems based on pre-trained LMs and has provided (semi-)automatic LM probing function which covers the tasks of term identification and inserting, taxonomy discovery, ontology alignment. OntoGPT is another Python package that uses GPT-3.5 and GPT-4 to extract and structure information from texts for various NLP tasks \cite{Caufield2023StructuredPI}. It offers two extraction strategies: SPIRES for recursive semantic structuring, and SPINDOCTOR for gene set summarization. Both of them support zero-shot learning and output in multiple data formats.

\label{6eval}

\section{Future Directions}
\label{direction}
In this section, we propose several noteworthy future directions, which we hope to inspire the readers for further exploration into the integration of LLMs with ontology learning tasks.

\subsection{Benchmarks Development}
Different from a single information extraction task like relation extraction or event extraction,  ontology learning is a procedure to construct an ontology, containing concept extraction, taxonomic relation extraction, non-taxonomic relation extraction and other steps. How to comprehensively analyze the quality of the constructed ontology is a problem worthy of study. The research on ontology learning using LLMs is at an initial stage, and the related research lacks  unified evaluation metrics and benchmarks. Advancements are anticipated in the field of the development of more sophisticated evaluation metrics to measure the performance of LLMs in ontology learning tasks, with the aim to develop benchmarks that can effectively measure the accuracy, completeness, and practical utility of ontologies generated by LLMs.

\subsection{Non-taxonomic Relation Extraction and Axiom Discovery}
At present, most of the existing researches on ontology learning using large language models only focus on taxonomy discovery. However, non-taxonomy relations and rules or axioms play an important role in ontologies, which improves the expressivity and explicity of ontologies. Many well-known ontologies contain abundant non-taxonomy relations and axioms \cite{yago, dbpedia}. Future work can lay emphysis on more sophisticated algorithms for non-taxonomy relations extraction and axioms formulation.

\subsection{Collaboration Between Domain Expert and Prompt Engineering}
We propose to investigate the utilization of interactive methodologies that involve domain experts in the knowledge acquisition process, as opposed to solely depending on prompting engineering. This type of collaboration has the potential to significantly improve the interpretive abilities of large language models. By incorporating domain experts and prompt engineering into the knowledge acquisition process, LLMs can be trained to capture domain-specific knowledge and context, enhancing their ability to interpret and generate meaningful responses. This approach may also lead to more accurate and contextually relevant language understanding, ultimately leading to better decision making and information processing capabilities.

\subsection{Leveraging LLMs for Dynamic Ontology Updating}
As knowledge in various fields continues to expand, the structure and content of ontologies become outdated, making it difficult for systems to accurately process and understand new knowledge. Therefore, it is essential to continuously update ontologies to reflect the latest knowledge and ensure that systems can maintain their performance and accuracy. Research is also warranted to examine the prospects of leveraging LLMs for dynamic ontology updating, enabling the systems to keep pace with the rapid evolution of knowledge domains.

\section{Conclusion}
\label{7con}

In this paper, we reviewed the progression of ontology learning from its inception, sketching the shallow learning and deep learning approaches. We focused on the advent of large language models for ontology learning. We highlighted how modern LLMs, like GPT-3.5 and GPT-4, have significantly advanced the field of ontology learning by enabling more sophisticated text interpretation and generation capabilities. These advancements pave the way for more automated, nuanced, and context-aware ontology learning processes.

For future work, we propose further exploration into the integration of LLMs with ontology learning tasks as follows: evaluation metrics and benchmarks development, non-taxonomic relation extraction and axiom discovery, collaboration between domain expert and prompt engineering, and leveraging LLMs for dynamic ontology updating.

%Bibliography
\bibliographystyle{unsrt}  
\bibliography{references}

\end{document}